\documentclass{aa}  
\usepackage{graphicx}
\usepackage{txfonts}
\def\nafe{[Na/Fe]}
\def\ffe{[F/Fe]}
\def\ofe{[O/Fe]}
\def\feh{[Fe/H]}
\def\alfe{[Al/Fe]}
\def\mgfe{[Mg/Fe]}
\def\ne22{$^{22}$Ne}
\def\tbce{T_{\rm bce}}
\def\msun{$M_{\odot}$}

\begin{document}
  \title{The self-enrichment scenario in intermediate metallicity
         globular clusters}
\authorrunning {Ventura \& D'Antona}
\titlerunning {AGB models of intermediate metallicity}


   \author{P. Ventura \and F. D'Antona}

   \offprints{P. Ventura}

   \institute{INAF - Observatory of Rome, Via Frascati 33,
              00040 MontePorzio Catone (RM) - Italy
              \email{ventura, dantona@oa-roma.inaf.it}
             }


 
  \abstract
   {We present stellar yields computed from detailed models of intermediate 
    mass asymptotic giant branch stars of low metallicity. In this work, the 
    whole main microphysics inputs have been updated, and in particular 
    $\alpha-$enhancement is explicitly taken into account both in the
    opacities and equation of state.}
   {The target of this work is to provide a basis to test the reliability of the
    AGB self-enrichment scenario for Globular Clusters of intermediate metallicity.
    These Globular Clusters exhibit well defined abundance patterns, which have 
    often been interpreted as a consequence of the pollution of the interstellar
    medium by the ejecta of massive AGBs.}
   {We calculated a grid of intermediate mass models with metallicity Z=0.001;
    the evolutionary sequences are followed from the 
    pre-Main sequence along the whole AGB phase. We focus our attention on
    those elements largely studied in the spectroscopic investigations of
    Globular Clusters stars, i.e. oxygen, sodium, aluminum, magnesium and fluorine.}
   {The predictions of our models show an encouraging agreement with the
    demand of the self-enrichment scenario for what concerns the
    abundances of oxygen, aluminum, fluorine and magnesium. The question 
    of sodium is more tricky, due to the large uncertainties of the
    cross-sections of the Ne-Na cycle. The present results show that only a 
    relatively small range of initial masses ($M=5,6$\msun) can be responsible 
    for the self enrichment.}
   {}


   \maketitle
%

\section{Introduction}
Deep spectroscopic investigations in the last decades {\bf have shown} 
that Globular Clusters (GC) stars are not chemically homogeneous 
samples, rather they show clear trends involving the chemical
abundances of some light elements (Kraft 1994), like magnesium,
aluminum, oxygen, fluorine and sodium; the same behaviour is not
followed by halo field stars, which present star to star variations limited
to carbon and nitrogen, i.e. those elements whose surface abundances 
are expected to change following the first dredge-up and the mixing 
possibly following the bump on the red giant branch (hereinafter RGB). 
In almost the totality of the GCs investigated, is present a main 
stellar population, whose surface chemistry is in agreement with the 
standard $\alpha-$enhanced abundances, and a second component, 
whose surface abundances of the afore mentioned elements define
well determined patterns (for a recent update, see Carretta 2006; 
Smith et al. 2005). These stars show depletion of oxygen and fluorine, while 
the abundances of sodium and aluminum are enhanced with respect to the 
solar values; the question of a  possible depletion of magnesium by 
$\sim 0.2$ dex is still under debate (Cohen \& Melendez 2005). A common 
feature of all the GCs investigated is that the C+N+O sum is constant 
within a factor $\sim 2$ (Ivans et al. 1999; Cohen \& Melendez 2005).

Two explanations, possibly acting together, are currently
investigated to explain such chemical anomalies. a) the deep mixing
scenario (DPS): a large non canonical extra-mixing, while
the stars ascend the RGB, might push inwards the convective 
envelope to reach layers where a very advanced nucleosynthesis 
(full CNO burning) might have acted (Denissenkov \& Weiss 2001;
Denissenkov \& Vandenbergh 2003; Denissenkov et al. 1998); 
b) the self-enrichment scenario (SES): an early generation of massive AGBs, 
evolving in the GC, eject into the interstellar medium material which is 
nuclearly processed, stimulating the formation of a second generation of 
stars, whose chemistry would trace the composition of the medium from which 
they formed (Cottrell \& Da Costa 1981; Ventura et al. 2001)
\footnote{Here we do not consider the alternative SES proposed by Maeder
\& Meynet (2006), Prantzos \& Charbonnel (2006) and described by
Decressin et al. (2007), namely that self-enrichment is produced by the
envelopes of fast rotating massive stars.}.

The capability of the DPS to fully explain the observed abundance patterns
was seriously undermined by the detection of the same chemical anomalies
even in main sequence and sub-giant stars of some GCs (Gratton et al. 2001), 
because the interior of these stars is not expected to reach temperatures 
so high to ignite a very advanced nucleosynthesis.

The question concerning the SES is still open, basically because many ingredients
used in the stellar evolution theory, whose physical formulation is not
directly based on first principles
(e.g. convection, mass loss rate, extra-mixing), have a strong impact on the
physical and chemical aspects of the AGB evolution (Ventura \& D'Antona 2005a;b); 
also, for one of the elements involved in the observed trends, i.e. sodium, 
the uncertainties associated to the relevant cross-sections ($\sim 3$ dex, Hale et al. 
2002; 2004) are such to render the results obtained very uncertain (Ventura \&
D'Antona 2006). Within the context of the Mixing Length Theory (MLT) modeling
of turbulent convection (Vitense 1953), the theoretical ejecta of AGBs 
point against the SES hypothesis, because the great number of third dredge 
up (TDU) episodes, associated with a modest nucleosynthesis achieved at the 
bottom of the outer convective envelope, favour a large growth of the 
overall C+N+O abundances, and inhibits the depletion of oxygen (Fenner et 
al. 2004; Denissenkov \& Herwig 2003).
On the other hand, Ventura \& D'Antona (2005b), studying the effects of 
changing the treatment of convection on the AGB modeling, show that when 
the Full Spectrum of Turbulence (FST, Canuto \& Mazzitelli 1991) is used to 
model the convective regions, the FST high efficiency 
of convection favours {\bf higher} temperatures at the bottom of the convective 
envelope, thus an advanced nucleosynthesis (which is usually referred to 
as ``Hot Bottom Burning'', HBB), a larger luminosity, and consequently a 
smaller number of TDU episodes, thus keeping the C+N+O sum almost constant.

An appealing prediction of the SES, and more generally of the role which
AGBs may play in the pollution of the interstellar medium within GCs, 
is that their yields are expected to be helium rich, as a consequence of
the deep second dredge up (hereinafter II DUP) experienced 
particularly by the most massive 
models. A helium content Y$>0.30$, as found in previous investigations
(Ventura et al. 2001), may be at least partially consistent with the existence
of helium rich stellar populations in GCs, which allow the interpretation
of the morphology of extended horizontal branches in some GCs (D'Antona et
al. 2002; D'Antona \& Caloi 2004; Lee et al. 2005); the presence of a
helium rich population was also invoked to explain the presence of a blue
main sequence in NGC 2808 (D'Antona et al. 2005b; Piotto et al. 2007) 
and $\omega$Cen (Bedin et al. 2004; Piotto et al. 2005).

The possibility that DPS and SES might act together in order to explain the
O-Na anticorrelation observed in a few giants of M13 has
recently been explored by D'Antona \& Ventura (2007), and was previously 
suggested by Denissenkov et al. (1998) to account for the Mg-Al 
trend in the same cluster.

{\bf Any SES model faces with a serious problem: it is difficult to understand how is
it possible that the self--enriched population is very abundant in most of the examined
clusters ---see, e.g., \cite{dantona-caloi2007}, concerning the ratio of the normal to 
self--enriched populations derived from the analysis of the Horizontal Branch (HB) 
morphology. In particular, for NGC~2808, both the HB ---\cite{franca1, dantona2005}---
the main sequence splittings ---\cite{dantona2005, piotto2007}--- and the spectroscopic
evidence ---\cite{carretta2006}--- indicate that about half of the cluster stars 
is self--enriched. In order to accomplish this we need that: 1) either the 
initial mass function (IMF) of the first generation stars is highly 
anomalous, and is peaked at the intermediate mass
stars; 2) or the IMF is more or less normal, the initial cluster mass is much larger, 
by about a factor 10, than the final mass, and the stars of the first generation have been 
preferentially lost, as discussed, e.g., in \cite{franca1} and \cite{prantzos1}.
In addition, \cite{bekki2007} suggest that all GCs
formed within dwarf galaxies, so that the cluster formation may take
advantage of the infall from {\it all} the gas lost by massive AGBs evolving in the
galaxy and falling into the protocluster potential well. 
} 
 
In this paper we model the AGB evolution by releasing some of the
approximations of our previous works. We focus on an 
intermediate metallicity (\feh$\sim -1.3$). We postpone to forthcoming 
papers the analysis concerning more metal rich and metal poor 
composition. Our approach is as follows: we present and 
discuss our intermediate mass models with metallicity Z=0.001, which were 
calculated with the latest and most updated physical inputs. Consistently 
with the hypothesis
which we want to test here, we adopt an $\alpha-$enhanced mixture.
We compare the yields of the individual elements with the abundances
observed in the GCs having metallicity appropriate for the present
investigation, taking care to restrict our comparison to those stars
which are scarcely evolved, to rule out any possible contamination from
canonical or non-canonical extra-mixing which the stars may experience
during the RGB evolution.
The effects of the uncertainties related to mass loss and the nuclear
reaction rates of the Ne-Na cycle and the Mg-Al chain are also examined.


\section{The ATON stellar evolution code}
The stellar evolution code used in this work is ATON3.1; a full
description of the numerical structure can be found in Ventura 
et al. (1998). Here we briefly recall the main physics inputs, with
the most recent updates.

\subsection{Opacities}
We adopt the latest opacities by Alexander \& Ferguson (2005) at temperatures 
smaller than 10000 K and the OPAL opacities in the version documented by Iglesias 
\& Rogers (1996). For both the above treatments we have the choice between the 
opacities corresponding to solar mixture (Grevesse \& Sauval 1998), and to an
$\alpha-$enhanced mixture $[\alpha$/Fe$]=0.4$. The conductive opacities are taken 
from Poteckhin (2006, see the web page www.ioffe.rssi.ru/astro/conduct/), 
and are {\bf harmonically} added to the radiative opacities.

\subsection{Equation of state}
Tables of the equation of state are generated in the (gas) pressure-temperature
plane, according to the method described in D'Antona et al. (2005a). For
mixtures including hydrogen, we use the latest OPAL EOS (2005), overwritten in
the pressure ionization regime by the EOS by Saumon, Chabrier \& Van Horn (1995).
The EOS is extended to the high density, high temperature domain according to the
treatment by Stoltzmann \& Bl\"ocker (2000). 
After these large tables are written, for a given Z , six values of
each physical quantity are computed for six different X values. A
cubic unidimensional spline provides the interpolation for any input
value of X. The six tables for H/He and given Z are supplemented
by 15 tables of He/C/O in which the EOS is directly computed
according to Stolzmann \& Blocker (2000) as the non-ionized regions
are not present in stellar structure following helium ignition. The
interpolation among the 15 tables is performed using triangles in
the plane C/O, as the stechiometric condition is $Y=1-X_C-X_O$.

\subsection{Convection}
The thermodynamic description of the regions unstable to convective
motions can be addressed either within the context of the traditional
MLT formulation (Vitense 1953), or by the FST model (Canuto et al. 1996). 

\subsection{Mixing}
Mixing of chemicals within convective zones can be addressed within
the instantaneous mixing framework or by a diffusive approach. In this
case, for each chemical species a diffusive-like equation 
(Cloutman \& Eoll 1976) is solved:
$$
{dX_i\over dt}=\big( {\partial X_i\over \partial t} \big)_{nucl}+
{\partial \over \partial m_r} \big[ (4\pi r^2\rho )^2D{\partial X_i \over \partial m_r} \big]
\eqno{(1)}
$$
where D is the diffusion coefficient, for which, given the convective velocity
$v$ and the scale of mixing $l$, a local approximation ($D\sim {1\over 3}vl$) 
is adopted.

The borders of the convective regions are fixed according to the
Schwarzschild criterium. It is also possible to consider 
extra mixing, by allowing convective velocities to decay exponentially
from the formal border, with an e-folding distance described by the
free-parameter $\zeta$ (see Ventura et al. (1998) for a complete
discussion regarding the variation of convective velocities in the 
proximities of the convective borders).

\subsection{Mass loss}
Mass loss can be treated according to different prescriptions. It is
possible to adopt the classic Reimers' treatment, the Vassiliadis
\& Wood (1993) formulation, or the prescription given by
Bl\"ocker (1995). In this latter case the strong increase of the mass
loss rate during the AGB evolution is modeled by multiplying the
Reimers' rate by an ad hoc luminosity power: the final expression is
$$
\dot M=4.83 \times 10^{-22} \eta_R M^{-3.1}L^{3.7}R
\eqno{(2)}
$$
where $\eta_R$ is the free parameter entering the Reimers' prescription.

\subsection{Nuclear network}
The nuclear network includes 30 elements (up to $^{31}$P) and 64
reactions. The full list of the 30 chemicals and of the reactions included
can be found in Ventura \& D'Antona (2005a).

The relevant cross section are taken from the NACRE compilation
(Angulo et al. 1999), with only the following exceptions:

\begin{enumerate}

\item{
$^{14}$N(p,$\gamma$)$^{15}$O (Formicola et al. 2004)
}

\item{
$^{22}$Ne(p,$\gamma$)$^{23}$Na (Hale et al. 2002)
}

\item{
$^{23}$Na(p,$\gamma$)$^{24}$Mg (Hale et al. 2004)
}
\item{
$^{23}$Na(p,$\alpha$)$^{20}$Ne (Hale et al. 2004)
}

\end{enumerate}

\section{The results of the most recent spectroscopic surveys of
intermediate metallicity GC}
Several clusters of intermediate metallicity (i.e. $-1.5 \leq$ \feh $\leq -1.2$)
have been investigated spectroscopically in the last decade: among these, the most 
extensively studied are M3,M4,M5,M13,NGC6752 and NGC6218. The first analyses
were focused on bright giants, and only in past few years turn-off (TO) and
sub-giant branch (SGB) stars were studied. Bright giants are of little help
for the main target of this paper, since the
abundances observed might be the result of some mixing episode from the bottom
of the convective envelope taking place during the RGB evolution; on the
other hand, TO and SGB stars never reach internal temperatures sufficiently high
to trigger any advanced nuclesynthesis, so any chemical anomaly must have
been imprinted in the matter from which they formed. Studies focused on the
TO and SGB stars are therefore the most useful for the present investigation.

In the analysis of the observational results, we will consider the uncertainties
associated to the data, and the constant offsets among
the different observers, mainly due to differences in the assumed solar
abundances and in the temperature scale.

The oxygen-sodium and magnesium-aluminum anticorrelations are by far the 
most studied trends. For NGC 6752, the works related to TO and SGB stars 
(Gratton et al. 2001; Carretta et al. 2005) and to bump stars (Gratton et 
al. 2005), evidentiated a well defined stellar population in which the oxygen 
was depleted up to 0.8 dex  (\ofe $\sim -0.4$) compared to the ``standard'' 
abundance (\ofe $\sim +0.4$), while sodium was enhanced at most by 0.4 dex. 
In Gratton et al. (2001) it is also shown that aluminum might be 
anticorrelated with magnesium, the most peculiar stars showing an aluminum 
enhancement slightly lower than 1 dex, and a magnesium spread of $\sim 0.3$dex.
For this cluster Pasquini et al. (2005) studied the lithium content of
TO stars, showing a clear anticorrelation between
sodium and lithium, the sodium rich stars having a lithium content 
A(Li)=2\footnote{We use the notation A(Li)=$\log($Li/H)+12 (in number)}, 
a factor of 2 smaller than the value detected in the stars with standard sodium.

The O-Na anticorrelation was recently confirmed
in the investigation of NGC 6218 performed by the same group 
(Carretta et al. 2006): for the stars below the RGB bump, the authors find a 
maximum extent of the oxygen depletion by $\sim 1$dex, and a maximum sodium 
enhancement of 0.5 dex.

Ramirez \& Cohen (2003) analyzed spectroscopically stars belonging to M5.
For the stars near the TO, they found a stellar population where sodium was
enhanced up to \nafe =0.3 dex, while for oxygen they could only find upper limits
for the estimated abundances: this is the main reason why the authors could not
detect any clear Na-O anticorrelation.

The case of M4 is hard to analyse, as only red giants spectra are available.
A first extensive exploration by Ivans et al. (1999) evidentiated an anticorrelation
among the CN weak stars, with a depletion of oxygen by $\delta$ \ofe $\sim 0.6$dex,
and an enhancement in sodium up to \nafe=0.3 dex. The most oxygen poor stars
in this subsample are aluminum rich (\alfe=0.8 dex). The magnesium abundance
is constant within 0.2 dex. A more recent work by Smith et al. (2005), unfortunately
restricted to giants, confirmed the extent of the oxygen depletion (keeping into
account a clear constant offset, which may be understood in the left-upper panel
of their fig.4) and of the aluminum enhancement found by Ivans et al. (1999);
they also found a clear anticorrelation between sodium and fluorine, the most
sodium rich stars being depleted in fluorine by [F/Fe]=-0.8 dex.

M3 and M13 have been extensively studied in the literature. 
Sneden et al. (2004) evidentiated a clear O-Na anticorrelation, which, for the
high gravity stars (i.e. those well below the RGB bump) extends to
\ofe$\sim -0.4$, and \nafe=+0.4. They also confirm, even for this
high gravity population, that the most oxygen poor and sodium rich stars 
are largely aluminum enhanced, with \alfe=1. The authors also claim the
detection in both clusters of a magnesium - aluminum anticorrelation, but this
result has been recently argued by Cohen \& Melendez (2005), who confirm
the extent of the oxygen depletion found by Sneden et al. (2004), but also
limit the sodium enhancement for the stars below the RGB bump to \nafe=+0.3.

%
\begin{table*}
\caption{Evolutionary properties of intermediate mass models}             
\label{tabfis}      
\centering          
\begin{tabular}{c c c c c c c c c c}     
\hline\hline       
$M/M_{\odot}$ & $\tau_{\rm H}/10^6$ & 
$\tau_{\rm He}/10^6$ & $\tau_{\rm AGB}/10^3$ & $\delta(M_{\rm 2dup})$ &
$M_{\rm core}/M_{\odot}$ & $\log(L/L_{\odot})_{\rm max}$ & 
$T^{\rm bce}_{\rm max}$ &  $N_{\rm pulse}$(TDU) & $\lambda$  \\ 
\hline                    
   3.0 & 275 & 49  & 410 & 0.01 & 0.76 & 4.40 & 74  & 6  & 0.7  \\  
   3.5 & 195 & 31  & 340 & 0.05 & 0.80 & 4.50 & 83  & 6  & 0.7  \\
   4.0 & 145 & 21  & 250 & 0.15 & 0.83 & 4.60 & 88  & 8  & 0.7  \\
   4.5 & 112 & 16  & 190 & 0.27 & 0.86 & 4.67 & 95  & 9  & 0.6  \\
   5.0 & 91  & 12  & 120 & 0.36 & 0.90 & 4.74 & 100 & 13 & 0.5  \\
   5.5 & 75  & 9.5 & 85  & 0.47 & 0.94 & 4.80 & 110 & 17 & 0.5  \\
   6.0 & 63  & 7.7 & 42  & 0.56 & 1.00 & 4.90 & 110 & 20 & 0.3  \\
   6.3 & 57  & 7.0 & 32  & 0.62 & 1.03 & 4.95 & 112 & 21 & 0.2  \\
\hline                 
\end{tabular}
\end{table*}

On the basis of the results quoted above, we may summarize the following
chemical features which characterize the stars in the GCs
which should have not experienced any ``in situ''
mechanism to change their surface abundances:

\begin{enumerate}

\item{The oxygen depletion is at most by $\delta [O/Fe] \sim 0.8$}

\item{The most aluminum rich stars show up an $^{27}$Al abundance of
\alfe=1}

\item{Sodium is anticorrelated to oxygen. The extent of the largest
sodium enhancement is \nafe=0.4, but this result is uncertain by 
at least 0.1 dex}

\end{enumerate}

Also, we may mention other two points, which must still be confirmed by
detailed spectroscopic investigations:

\begin{enumerate}

\item{Magnesium might be correlated to oxygen, and anticorrelated to
sodium and aluminum}

\item{Fluorine might be strongly anticorrelated to sodium, but this
result was found at the moment only for giant stars}

\item{Lithium is correlated to oxygen: the oxygen poor population
shows a lithium content smaller by a factor 2}

\end{enumerate}

\section{The physical properties of the intermediate mass models}
The models were evolved from the pre-MS through the whole AGB phase.
The chemistry adopted is typical of the intermediate metallicity 
GCs, i.e. Z=0.001 and Y=0.24.
The mixture, consistently with the main target of this investigation, is
assumed to be $\alpha-$enhanced, with [$\alpha$/Fe]=+0.4; the individual
abundances are taken from Grevesse \& Sauval (1998).
The range of masses involved is $3M_{\odot}\leq M \leq 6.3M_{\odot}$; the
limits were chosen to restrict the analysis to stars achieving HBB
during the AGB phase, and not undergoing any carbon ignition in the 
interior.
Convection was modelled according to the FST treatment. 
Due to the importance that CNO burning within the most
internal part of the convective envelope may have on the AGB evolution
(Herwig 2005; Mazzitelli et al 1999),
we adopted the diffusive treatment in all the models presented
here. In the phases preceeding the AGB evolution, a free 
parameter $\zeta=0.02$ was used to model the exponential decay of velocities 
within regions stable against convection;
this is in agreement with the calibration in Ventura et al. (1998).
For the whole AGB phase no extra-mixing was assumed from any convective
border. Mass loss was modelled according to Bl\"ocker (1995), with the free 
parameter entering eq.2 set to $\eta_R=0.02$, according to the
calibration given in Ventura et al. (2000).

\begin{figure*}
\centering{
}
\caption{Variation with the total mass of the luminosity (Left) and
Temperature at the bottom of the convective envelope (Right) of some
intermediate mass models during the AGB phase. For clarity reasons, only
models corresponding to initial masses 3,4,5,6$M_{\odot}$ are shown.}
      \label{agbfis}%
\end{figure*}

The main physical properties of the models are reported in
Table \ref{tabfis}. Cols. 2 to 4 show the duration of the core H- and He-
burning phases (Myr) and of the AGB phase (Kyr). 

Within stars of intermediate mass, after the extinction of the CNO 
burning shell following the core He-burning phase, the bottom of the convective
envelope sinks inwards, down to layers precedently touched by
nuclear burning; this is the second dredge-up (Iben 1991)
Col.5 reports the total mass previously involved in CNO burning
mixed with the convective envelope during the II DUP; in the most
massive models, where the amount of dredged-up matter is higher, 
the surface chemistry is altered, with
a decrease of the surface oxygen abundance and an increase of the sodium
and helium mass fractions.

These results are consistent with the old models calculated with the same
metallicity, presented in Ventura \& D'Antona (2005b, see Table 1).
In cols. 6-10 of Table \ref{tabfis} we give some details of the AGB evolution,
namely the core mass at the beginning of the Thermal Pulses (TPs) phase,
the maximum luminosity and the maximum temperature reached at the bottom 
of the convective zone ($\tbce$), the number of the first TP followed by 
a TDU episode, and the maximum efficiency of the TDU during the whole 
stellar AGB life\footnote{We use here the usual descriprion of the
efficiency of the TDU in the terms of the quantity $\lambda$, defined
as the ratio of the total matter dredge-up after a TP to the
mass by which the H-exhausted core advanced from the previous TP}.

Fig.\ref{agbfis} shows the variation with mass (decreasing
during the evolution) of the luminosity and $\tbce$ in some of our models. 
We note in all cases, with
the only possible exception of the $3M_{\odot}$ model, a rapid increase
of the luminosity after the very first TPs, associated to an increase
of $\tbce$. This is a consequence of the high efficiency of the convective
model adopted, as correctly predicted by Bl\"ocker \& Sch\"onberner (1991),
and later confirmed by D'Antona \& Mazzitelli (1996) and 
Ventura et al. (2000). A detailed comparison of the results obtained with
various efficiencies of the convective model can be found in Ventura
\& D'Antona (2005b). 

The large luminosities attained by our models have two important
consequences:

\begin{enumerate}

\item{A fast decrease of the mass of the envelope, with a consequent
small number of TPs, and therefore of TDUs}

\item{A very advanced nucleosynthesis at the bottom of the envelope}

\end{enumerate}

In the most massive models, mass loss is so large that they  
reach their maximum luminosity after a few TPs; they loose their 
envelope so rapidly, that TDU takes place only in the latest stages of 
their AGB evolution, and with a very modest efficiency 
(see col.10 of Table \ref{tabfis}).
We stress that the larger is the luminosity, the faster is the 
general cooling of the structure determined by the gradual loss of the
mass of the envelope: the maximum temperature achieved at the bottom
of the external convective zone reaches a maximum asymptotic value
(see col.8 of Table \ref{tabfis}), which, for large M, turns out to be 
independent of the total initial mass of the star; in the present 
computations, this upper limit is $\tbce=110$MK.

%
\begin{table*}
\caption{Chemical composition of the ejecta of intermediate mass models}             
\label{tabchim}      
\centering          
\begin{tabular}{c c c c c c c c c c c c}     
\hline\hline       
$M/M_{\odot}$ & $\eta_R$ & Y & A(Li) & [$^{12}$C/Fe] & [$^{14}$N/Fe] &
[$^{16}$O/Fe] & [$^{19}$F/Fe] & [$^{23}$Na/Fe] & [Mg/Fe] & [$^{27}$Al/Fe] & 
R(CNO) \\ 
\hline                    
   3.0 & 0.02 & .248  & 2.77  & 0.84   & 2.21  &  0.92  &   0.10  &  1.16  &  0.57  &  0.65  & 9.6    \\  
   3.5 & 0.02 & .265  & 2.43  & 0.51   & 2.18  &  0.77  &  -0.26  &  1.30  &  0.55  &  0.66  & 7.9   \\
   4.0 & 0.02 & .281  & 2.20  & 0.14   & 2.02  &  0.44  &  -0.61  &  1.18  &  0.48  &  0.55  & 4.9   \\
   4.5 & 0.02 & .310  & 2.00  & 0.12   & 1.89  &  0.19  &  -0.90  &  0.97  &  0.43  &  0.85  & 3.1   \\
   5.0 & 0.02 & .324  & 1.98  & 0.13   & 1.70  & -0.06  &  -1.16  &  0.60  &  0.35  &  1.02  & 2.1   \\
   5.5 & 0.02 & .334  & 1.93  & -0.41  & 1.51  & -0.35  &  -1.39  &  0.37  &  0.28  &  1.10  & 1.3   \\
   6.0 & 0.02 & .343  & 2.02  & -0.62  & 1.35  & -0.40  &  -1.36  &  0.31  &  0.29  &  1.04  & 0.97  \\
   6.3 & 0.02 & .348  & 2.06  & -0.68  & 1.33  & -0.37  &  -1.28  &  0.30  &  0.30  &  0.99  & 0.94  \\
   5.0 & 0.01 & .327  & 1.79  &  0.00  & 1.83  & -0.14  &  -1.30  &  0.67  &  0.40  &  1.20  & 2.77  \\ 
   5.0 & 0.04 & .323  & 2.49  & -0.37  & 1.58  & -0.05  &  -1.00  &  0.70  &  0.39  &  0.80  & 1.71  \\
   6.0 & 0.01 & .345  & 1.83  & -0.40  & 1.49  & -0.42  &  -1.45  &  0.27  &  0.22  &  1.07  & 1.29  \\
\hline                 
\end{tabular}
\end{table*}

\section{The chemistry of the ejecta}
For each model, we calculate the average mass fractions in the ejecta, 
for the chemical elements included in our network. 
The results for the species of interest for this work, are presented in 
Table \ref{tabchim}. For any isotope A, we give the quantity [A/Fe], defined 
as [A/Fe$]=\log($A/Fe$)_{\rm ejecta}-\log($A/Fe$)_{\odot}$; the abundances are 
mass fractions. The only exceptions are helium and lithium, for
which we list, respectively, the mass fraction Y and the standard A(Li) quantity.
Note that Mg in col.9 refers to the total magnesium abundance, and R(CNO) in the 
last column gives the ratio between the total C+N+O in the ejecta and the initial
value.

\begin{figure*}
\centering{
}
\caption{Left: Variation with the total mass of the surface oxygen mass
fraction of the AGB models with masses $3M_{\odot} \leq M \leq 6M_{\odot}$;
the strong depletion in the most massive models is a clear signature of
strong HBB at the bottom of the convective envelope. Right: oxygen 
content of the ejecta as a function of the initial mass; a plateau value
of \ofe=-0.4 is reached for the highest masses}
      \label{oxy}%
\end{figure*}

\subsection{The oxygen depletion}
The depletion of the surface oxygen abundance requires a strong HBB at the
bottom of the convective envelope, as the activation of the full CNO cycle
demands temperatures approaching 100MK, which, within the context of the
AGB modelling, are attainable only via a very efficient description of
convection (Ventura \& D'Antona 2005b). The present models have been
calculated with the FST prescription, thus, at least in the most massive
models, we expect to reach such a high $\tbce$, as can be seen in the right
panel of Fig.\ref{agbfis}.

The left panel of Fig.\ref{oxy} shows the variation with the total mass
of the surface oxygen abundance for the masses examined here, with the
only exception of the 6.3\msun model, which was omitted for clarity reasons,
being very similar to the 6\msun  case. 
We see that oxygen burning starts efficiently shortly after the beginning of
the AGB phase in all the models more massive than 4\msun. In the same panel
we may easily note the effects of the TDU, which, when sufficiently 
efficient, increases the oxygen content of the envelope, because carbon and
oxygen rich material is dredged-up from the ashes of the precedent $3\alpha$
burning shell which forms during each TP. A strong oxygen depletion is thus
inhibited by repeated and efficient TDU episodes, so that in the less massive
models, which undergo many TDUs, and hardly reach the temperatures requested
to ignite oxygen burning, the oxygen is indeed produced rather than destroyed
within their envelopes.
With increasing mass, we shift progressively to a situation where the final
oxygen abundance is a delicate compromise between the depletion triggered by
the activation of the full CNO cycle and the increase due to the TDU
(4.5\msun$\leq M \leq$ 5.5\msun), to end up with the most massive models
($M\geq$6\msun), in which oxygen can be eventually depleted by a factor 
$\sim 20$ compared to the initial value (see the track corresponding to 
the 6\msun model in the left panel of Fig.\ref{oxy}).

In the right panel of Fig.\ref{oxy} we show the oxygen content of the ejecta 
of the models calculated, in terms of \ofe, to allow a more straight comparison
with the observations outlined in Sect.2. We note a very high oxygen content 
in the ejecta of the models with masses $M<4$\msun, consistently with the
previous discussion; the model with initial mass 4\msun  shows an
oxygen content unchanged compared to the initial $\alpha-$enhanced value.
The mass-oxygen trend is progressively decreasing with mass, and reaches a
plateau value of \ofe$ \sim -0.4$dex for all the masses $M>5$\msun. The reasons
why this lower limit exists is twofold: a) on the one hand (see col.8 in 
Table \ref{tabfis}), we saw that there is an upper limit for $\tbce$,
which therefore limits the degree of O-burning which may be achieved at
the bottom of the envelope; b) the luminosity in the most massive models is
so high that they loose mass rapidly, already from the very first TPs, when
the oxygen abundance is still large (see the different slope of the Oxygen-Mass
relation characterizing the 6\msun model compared to the other masses
in the left panel of Fig.\ref{oxy}). We will show that changing the mass loss
description does not change substantially this conclusion.
Finally, we note that this lower limit for \ofe is in good agreement with the
lowest oxygen abundances measured in TO and SGB stars belonging to intermediate 
metallicity GCs.

\subsection{Aluminum production and the activation of the Mg-Al cycle}
Cols. 10 and 11 of Table \ref{tabchim} report the Mg and Al content of the
ejecta.
While the observed spread in the Mg abundance, as outlined in Sect.3, is still
debated, and in any case restricted to 0.2-0.4 dex, the stars in GCs showing 
up the largest degree of oxygen depletion are also strongly enriched in 
aluminum, with \alfe=1 in the most extreme cases (Gratton et al.2001;
Sneden et al. 2004).

In the AGB modelling, an aluminum production at the bottom of the envelope
is made possible by the activation of the Mg-Al chain, which, favouring 
proton captures by the heavy isotopes of magnesium, eventually leads 
to the synthesis of $^{27}$Al (Denissenkov \& Herwig 2003; Ventura \&
D'Antona 2005a). Also TDU plays a role which is not negligible, 
as a deep penetration 
of the convective envelope within the region precedently touched by helium
burning may bring to the surface $^{25}$Mg and $^{26}$Mg synthetized via 
$\alpha-$ captures by $^{22}$Ne nuclei; these isotopes, once the CNO burning 
shell is reactivated, undergo proton capture and produce aluminum.

\begin{figure}
\caption{The variation during the AGB phase of the surface aluminum
abundance, for the same models reported in fig.\ref{oxy}. Note the combined
effects of HBB and of TDU to increase the surface aluminum content}
      \label{al}%
\end{figure}

Fig.\ref{al} shows the evolution of the surface aluminum content in the
models presented here. We note in all cases a trend increasing during the
evolution; in the most massive models, due to the stronger
nucleosynthesis activated, the aluminum production is larger.
Even in this case, as it was for the depletion of oxygen, we note an
upper limit to the aluminum enhancement, which can also be seen in
col.10 of Table \ref{tabchim} to be \alfe=1. The reasons for the
impossibility of a larger Al production are the upper limit of $\tbce=110$MK
found in our models, the negligible extent of TDU in the most massive stars,
preventing the transport of the magnesium isotopes from the interior to
the surface, and the strong mass loss suffered from these models at the
beginning of the AGB phase, which favours the ejection into the ISM
of material which is not aluminum rich.

We see from col.11 of Table \ref{tabchim} that the ejecta of all the models
are characterized by $0.5\leq$\alfe$\leq 1$, and are
therefore fully consistent with the aluminum content of the stars with
the anomalous chemistry outlined by the spectrospic investigations 
presented in Sect.3. 

The total magnesium abundance, as it is evident from the previous discussion,
is the result of the balance between the increase of Mg determined by the 
TDU, and the depletion due to proton captures during HBB. This explains
the negative trend with mass which can be seen in col.9 of Table \ref{tabchim}.

We underline here the striking difference between our findings and the
results obtained by Fenner et al. (2004, see the bottom panel of their
fig.1), where they found that the most Al-rich stars were also magnesium 
rich. This was a result of the effects
of many TDUs, enriching the envelope with the heavy magnesium isotopes
produced in the $3\alpha$ shell; it is the different treatment of convection
between the two sets of models leading to this discrepancy, because the
use of the FST model reduces the number of TPs and TDUs.

Contrary to oxygen, the predictive power of our results {\bf for Al}
is undermined by the range of uncertainties related to the relevant cross-sections.
Izzard et al. (2007) evidentiated that in massive AGBs the
yields of $^{27}$Al is affected by the uncertainties connected to both the
$^{26}$Mg(p,$\gamma)^{27}$Al and $^{26}$Al(p,$\gamma)^{27}$Si reaction rates: 
we used
the upper NACRE limit for these reactions in the present investigation, but we
keep in mind that these results have an associated uncertainty which may be
estimated to be around 0.3-0.4 dex. 

\subsection{The puzzling behaviour of sodium}
The debate regarding the amount of sodium which may be synthesized within
AGBs is still open, due to the large uncertainties associated to a) 
the cross-sections of the reactions involved in the Ne-Na cycle; b) the
cross sections of the $\alpha-$captures by \ne22;
c) other physical inputs which play a role in determining the sodium 
content within the envelope of these stars.

As outlined by Ventura \& D'Antona (2006), the surface sodium abundance first increases
due to the II DUP, then, particularly in AGB models calculated with an 
efficient convective model, it is further produced by burning of the
dredged-up $^{22}$Ne, 
and is later decreased when the rate of destruction exceeds that of production.
Any TDU favours sodium production, due to dredging up of primary $^{22}$Ne 
synthesized
via $\alpha-$capture within the convective shell which forms during the TP.
This behaviour is confirmed by Fig.\ref{sodio}, which shows the variation of
the surface sodium abundance of the evolutionary models.
The behaviour of the M$ \leq 4$\msun  stars is in qualitative agreement with the AGB
models used by Fenner et al.(2004) (see the upper panel of their Fig.1): 
we note a great increase of the sodium abundance, due to
the dredge-up of $^{22}$Ne which is later converted to sodium.
Contrary to their findings, our more massive models show an opposite behaviour, 
because the FST convective model favours a more advanced nucleosynthesis, with
a partial destruction of the sodium precedently created; also, we recall that
the higher mass loss favours a smaller number of TPs, thus acting against
sodium production.

\begin{figure}
\caption{Variation with mass of the surface sodium abundance within
our standard models}
      \label{sodio}%
\end{figure}

\begin{figure}
\caption{Variation of the sodium surface content within models with
initial masses 4,5,6\msun calculated with the recommended values of
the cross-sections of the Ne-Na cycle (dotted), and those maximising
sodium production (see text for details). The dashed tracks show the
results obtained when the upper limits of the $\alpha-$capture
reactions by $^{22}$Ne nuclei are adopted.}
      \label{sodio2}%
\end{figure}

The precedent discussion explains the clearly negative trend with mass
of the \nafe\quad values in the 9th column of Table \ref{tabchim}.
We focus our attention on the most massive models, which produce yields
which we saw to be aluminum rich and oxygen poor: the sodium content
of their ejecta is in the range 0.3-0.4 dex, which is consistent with 
the sodium abundances derived by most of the research groups
for the stars with the most anomalous chemical composition.

The uncertainties connected to the cross sections have a dramatic impact
on the value of \nafe\quad of the ejecta, even more than we saw for aluminum.
Concerning HBB, the main problems are associated to the cross section of the 
$^{22}$Ne(p,$\gamma)^{23}$Na reaction, which is uncertain by a factor 
$\sim 2000$ (Hale et al. 2002); even the other reaction relevant to 
determine the correct sodium equilibrium value, i.e. 
$^{23}$Na(p,$\alpha)^{20}$Ne, has a margin of uncertainty, which is 
however smaller ($\sim 30\%$; Hale et al. 2004).

\begin{figure*}
\centering{
}
\caption{Left: Variation as a function of the total stellar mass of the
surface lithium content of the intermediate mass models during the whole
AGB evolution. Right: The lithium content of the ejecta, as a function of
the initial mass.}
      \label{litio}%
\end{figure*}

We {\bf ran} three models with initial masses 4,5,6\msun, where we used the 
lower limit for the $^{22}$Ne(p,$\gamma)^{23}$Na reaction, 
and the upper limit for $^{23}$Na(p,$\alpha)^{20}$Ne. The comparison 
between these simulations (dotted tracks) and those described in 
Table \ref{tabchim} (solid) {\bf is} shown in Fig.\ref{sodio2}.
We note that when the cross sections minimizing the sodium production are used, 
with the only exception of the II DUP, sodium is destroyed when $\tbce$ becomes 
sufficiently large to ignite proton capture by $^{23}$Na nuclei. The sodium contents
which we get are considerably lower, i.e. \nafe=0.6 for the 4\msun model,
\nafe=0.0 in the M=5\msun case, and \nafe=-0.2 for M=6\msun; 
these values are smaller than those
reported in Table \ref{tabchim} by 0.6 dex.
Note that this is not proportional to the reduction factor of the 
$^{22}$Ne(p,$\gamma)^{23}$Na reaction, because, as pointed out
by Izzard et al. (2007), once $^{22}$Ne is destroyed at the bottom of
the envelope, no further sodium can be created, despite the use of a cross
section for $^{22}$Ne burning which is a factor $\sim 2000$ higher.
This is also consistent with the investigation by Ventura \& D'Antona (2006),
who pointed out that a decrease of the $^{23}$Na(p,$\alpha)^{20}$Ne 
reaction by a factor 2 could reconcile better the sodium content of 
the ejecta of the most massive AGBs of intermediate metallicity with 
the spectrospic measurements of GC stars.

For the models experiencing TDU, another source of uncertainty
for the sodium yield is provided by the cross sections of the $\alpha-$ 
capture reactions by \ne22 nuclei, whose upper limit is 3 orders
of magnitude higher than the recommended values (Angulo et al. 1999).
These reactions determine the \ne22 content in the convective shell
which forms during the TP, hence the amount of \ne22 which may be dredged-up 
in the after-pulse phase, and therefore the quantity of sodium which may be 
sinthesized at the bottom of the envelope via proton capture by \ne22 
nuclei.
 
The dashed tracks in Fig.\ref{sodio2} show the results of the surface 
sodium abundance when these upper limits are used: for the 4 and 5\msun
the overall sodium abundance is clearly reduced.
Differently from the previous case, the associated uncertainties 
on the sodium yields are not uniform with mass, but rather show a
decreasing trend, ranging from a null effect for the most
massive models, to $\delta [Na/Fe]\sim 0.15$dex for $M=5M_{\odot}$, up to 
$\delta [Na/Fe]\sim 0.3$dex for $M=4M_{\odot}$; 
these results are actually not surprising, since dredging up of \ne22
becomes progressively more important in determining the sodium abundance 
the smaller is the stellar mass.

The yields of sodium from these sources are still highly uncertain; only
a more solid estimate of the relevant cross-sections may help increasing the
reliability of these investigations.

\subsection{The lithium problem}
Lithium is synthesized at the bottom of the convective envelope
of AGBs at temperatures exceeding $40 \times 10^6$K via the Cameron \&
Fowler (1971) mechanism. Sackmann \& Boothroyd (1992) first showed that
the use of a diffusive approach was mandatory to describe such a delicate
interplay among nuclear and mixing time scales, which could eventually
lead to lithium production. Ventura et al. (2000) used this approach to
reproduce the lithium vs. luminosity trend observed in the Large Magellanic
Cloud.

The key factor to achieve lithium production is the activation of
the $^3$He($\alpha,\gamma)^7$Be reaction at the bottom of the convective
envelope, which is possible when $\tbce > 40 \times 10^6$K. 

The left panel of Fig.\ref{litio} shows that all our models 
attain temperatures at the bottom
of the convective envelope sufficiently high to ignit the Cameron \& Fowler
mechanism; the most massive stars undergo a rapid consumpion of the whole
$^3$He available in the envelope, so that the surface lithium content, after
reaching a maximum value A(Li)$\sim 4$, rapidly declines to extremely low
abundances. This process becomes progressively slower as the mass decreases; we
end up with the 3\msun model, which is still lithium rich at the end of its 
AGB evolution. This discusion explains the relation between the
initial mass of the star and the lithium content of its ejecta, shown in the
right panel of Fig.\ref{litio}; all the massive models have
A(Li) $\sim 2$, in excellent agreement with the lithium abundance of the
oxygen poor TO stars in NGC 6752 (Pasquini et al. 2005).

\subsection{The helium enrichment}
The AGB models discussed here experience a small number of TPs, thus most of the 
helium enrichment of the envelope takes place during the II DUP. Since this latter
is deeper the higher is the initial mass of the star (see col.5 of Table \ref{tabfis})
we expect the helium enrichment to increase with mass, as it is confirmed by the
results reported in col.3 of Table \ref{tabchim}. We find a small increase (compared 
to the standard Big Bang value, Y=0.24) in the models with mass $M<4$\msun; the
maximum enrichment, for the masses close to the limit for carbon ignition, is
Y=0.35. We discuss in a forthcoming paper (Pumo et al. 2007) the overall helium 
enrichment due to AGB and super-AGB stars of intermediate metallicity, and the 
robustness of these predictions, related essentially to the efficiency of the 
II DUP.

\subsection{Comparison with previous models}
Ventura et al. (2002) presented models for the evolution of stars of
intermediate mass at various metallicities, ranging from 
$Z=2\times 10^{-4}$ to $Z=0.01$. We compare the yields of the present
work with the results of that investigation, for the metallicity
$Z=0.001$. First, we note that the helium content is sistematically
higher here by $\sim 0.04$, as can be seen by comparing the third
column of Table \ref{tabchim} with the dashed curve giving the helium-mass
trend in fig.4 in Ventura et al. (2002). This results may be understood in
terms of the extra-mixing from the bottom of the convective envelope which
was assumed in the present models, and which was neglected in Ventura et al. 
(2002).

The combined effects of the overshooting from the envelope and the
different mixture adopted (solar in Ventura et al. (2002), $\alpha-$
enhanced here), favours smaller core masses in the present models, so that
a given model here can be compared with a model $0.5M_{\odot}$ less
massive in Ventura et al. (2002). 

Even with this assumption, we note differences in the CNO yields, the
overall C+N+O contribution being higher here compared to the corresponding
values in Ventura et al. (2002) (see fig.6 in Ventura et al. 2002, to be
compared with columns 5 to 7 in Table \ref{tabchim}). The reason for such
discrepancy is the different efficiency of the TDU found in the two sets
of models: we find a maximum efficiency of $\lambda=0.7$ here (which decreases
to $\lambda=0.5$ in the more massive models), whereas in Ventura et al. (2002)
$\lambda$ could hardly reach 0.5. This change, favoured by the different mixture 
adopted, determines the differences in the yields obtained.  

The detailed comparison of the other yields is rendered hard by the differences
in the nuclear cross-sections adopted for the relevant reactions; Ventura et al. (2002)
used the NACRE compilation, whereas here we use the most updated releases present
in the literature.

\section{The uncertainties related to mass loss}
Mass loss plays a fundamental role in the context of AGB evolution: it is
the reduction of the convective envelope via stellar winds which eventually
halts the TP phase, and leads to PN ejection. Also, the description of mass loss
determines the number of TPs experienced by the star, and, in turn, the
number of TDUs (Sch\"onberner 1979).

Ventura et al.(2000) calibrated the parameter $\eta_R$ entering eq.(2) by
reproducing the luminosity function of lithium rich stars in the
Magellanic Clouds; the chemistry of the stars examined in the present work is
different, thus leaving some room for a possible variation of $\eta_R$. It is
therefore essential to understand to which extent the yields which we obtain 
depend on this choice.

On the physical side, a larger mass loss 
leads the star evolve at lower luminosities; the degree of the 
nucleosynthesis achieved at the bottom of the convective envelope is
reduced, because we have smaller temperatures.
This is confirmed by fig.\ref{mloss1}, where we show the evolution of the 
luminosity and of $\tbce$ in models with initial masses 5 and 6\msun, 
calculated with different $\eta_R$s.

Chemically, the situation is complex, and not all the isotopes
follow the same trend with $\dot M$;  this is
dependent on the modality with which any chemical species
is synthesized (or destroyed) within the convective envelope.
The last three lines of Table \ref{tabchim} report the chemistry of the
ejecta of the models of 5 and 6\msun calculated with a different mass loss
rate with respect to our standard case.

The behaviour of lithium is the most linear. As can be seen in the left panel
of fig.\ref{mloss2}, lithium is produced only during the first TPs, so that the
average lithium content of the ejecta is determined essentially by the mass
lost by the star during this phase. A large $\dot M$ allows a larger release
of lithium rich material during the early phases of the AGB evolution, with a
consequent increase of A(Li): for a given mass, we see in Table \ref{tabchim}
an almost linear trend A(Li)-$\eta_R$. A similar, straight path is also
followed by fluorine, which is destroyed during the very first TPs, so that 
the overall yield is determined by the strength of the stellar winds during
the first TPs; in this case the slope of the \ffe-$\eta_R$ relation is 
lower compared to lithium, because the fluorine consumption is faster 
than the duration of the whole phase of lithium production and destruction.

Other elements show up a less defined behaviour, because their yield is
determined by the nucleosynthesis at the bottom of the envelope during 
the whole AGB phase, and also by the TDU. Interestingly, we find that in
some cases the yields are not very sensitive to the mass loss rate adopted.
A typical example is oxygen, for which we show the
variation of the surface abundance in the right panel of fig.\ref{mloss2}.
When mass loss is reduced, the tendency of the oxygen
abundance of the ejecta to diminish (less oxygen-rich matter during the
first TPs is lost by the star) is partly compensated by the larger number
of TPs, which act to increase the surface oxygen content.
The clearest example is the 6\msun model, for which 
a smaller $\eta_R$ (dotted track) leads to a smaller surface oxygen 
abundance until the total mass of the star drops to 3\msun, but to a
larger mass fraction in the latest evolutionary phases, when the TDU
becomes efficient.
As can be seen in Table \ref{tabchim}, the oxygen content of the ejecta is almost 
independent of $\eta_R$, and so is the general conclusion which we reached
in Sect.5.1 concerning the maximum depletion of oxygen obtainable by these
models: playing with mass loss leaves unchanged the maximum depletion of
oxygen obtainable at these metallicites, leading to a minimum oxygen content
of the ejecta \ofe=-0.4.

A behaviour similar to oxygen is also followed by sodium, again because its 
abundance in the envelope is a balance between destruction via proton capture 
and production via dredging up of $^{22}$Ne. Independently of mass, the 
related uncertainty is $\delta$\nafe$\sim 0.1$dex.  

The aluminum abundance of the ejecta decreases with $\dot M$, because
a larger mass loss favours a larger release of mass at the beginning of the
AGB phase, when the aluminum has not yet been synthesized in great amounts.
The smaller number of TPs at large $\dot M$ tends to increase 
the $^{27}$Al abundance indirectly, via dredge-up of the heavy magnesium isotopes, 
which later form aluminum via proton capture. The variation of \alfe with the 
mass loss rate is ($\delta \alfe \sim 0.1$dex).

Helium is not influenced by the details of the mass loss description,
because the only {\bf substantial} change of its surface abundance takes place
during the II DUP, and remains approximately constant during the whole
following AGB phase.

Finally, we note from the last column of Table \ref{tabchim} that the CNO ratio
has a negative trend with mass loss, which is a mere consequence of the fact
that when $\dot M$ is high the star experience a smaller number of TPs.

We may summarize the effects of the uncertainties of mass loss during the
AGB phase on the chemistry of the ejecta as follows:

\begin{enumerate}

\item{Oxygen is not sensitive to the mass loss rate adopted}

\item{Sodium and aluminum show a positive trend with mass loss, but the
{\bf uncertainty associated with the adopted description of mass loss}
is considerably smaller than the indetermination due to the unknown 
relevant cross-sections}

\item{Lithium and fluorine are more sensitive to mass loss, with a
linear positive trend}

\item{The CNO content of the ejecta is sensitive to mass loss, because
a stronger $\dot M$ diminishes the number of TPs experienced by the
stars, thus favouring a smaller C+N+O abundance}

\end{enumerate}    

\begin{figure}
\caption{Comparison between the variation with the total mass of the
luminosity (top) and temperature at the bottom of the convective envelope
(bottom) of models with initial masses 5,6\msun, calculated with
different values of the parameter $\eta_R$ enetering eq.2}
      \label{mloss1}%
\end{figure}

\begin{figure*}
\centering{
}
\caption{Variation with the total mass of the star of the lithium (Left)
and oxygen (Right) surface abundances within models of initial mass
5 and 6\msun calculated with different mass loss rates. Note the
straightforward dependency of lithium on mass loss, compared to the
more tricky behaviour of oxygen, whose destruction is first amplified
by a lower mass loss rate, and later prevented by TDU episodes}
      \label{mloss2}%
\end{figure*}

\section{Discussion: theory vs. observations}
We discuss the self-enrichment scenario hypothesis for intermediate 
metallicity clusters, by asking whether the ejecta of the most massive 
AGBs can account for the chemical patterns traced by the abundances of 
the GC stars with the most anomalous chemistry. We restrict our attention 
on the least evolved stars (i.e. TO and SGB sources, 
or giants well below the RGB bump),
despite the difficulties presented by their spectroscpic analysis,
because this allows to rule out any possible change of the surface 
chemistry due to some non-canonical extra mixing while ascending the
RGB; we therefore disentangle the primordial from the evolutionary 
effects, and focus only on the abundance patterns present directly 
in the matter from which the stars formed.

Our goal is {\bf to} test the possibility that our ejecta can reproduce 
the observed O-Na and O-Al trends, and that the stars showing the
strongest oxygen depletion are fluorine poor and possibly sligthly
depleted in magnesium. Finally, we compare our results with the
recent analysis of the lithium abundances in NGC 6752, which 
indicate that oxygen poor stars deviate from the Spite's plateau,
having a lithium content a factor 2 smaller than the standard
value (Pasquini et al. 2005). 

In making this comparison, we keep in mind that the oxygen abundances
predictions are very robust, that the aluminum and even more sodium
mass fractions are made uncertain by the poor knowledge of the relevant
cross-sections in the range of temperatures of interest here (T$\sim 100$MK), 
and that lithium is strongly influenced by the assumed mass loss rate 
at the beginning of the AGB phase.
  
The left panel of fig.\ref{antic} shows the observed oxygen-sodium trend
for such stars. The dashed track and the two solid lines indicate the 
abundances of the ejecta of our models, differing in the chosen cross-sections 
for the reactions involving sodium: the upper solid track refers to models calculated 
with the upper limits for the $^{22}$Ne(p,$\gamma)^{23}$Na and the lower
limits for the $^{23}$Na(p,$\alpha)^{20}$Ne reactions, while the lower 
refers to models calculated with the recommended values for the same
reactions; the dashed track indicates the results obtained by maximizing the
rates of the $\alpha-$capture reactions by $^{22}$Ne.

According to our interpretation, the observed points
inside the squared box in the right lower portion of the plane represent stars
born with the original chemistry, while those belonging to the 2nd generation,
whose initial chemistry traces the pollution by AGBs, are included within the
squared box in the left-upper part. The remaining points, with high sodium
and normal oxygen, may be stars formed by processed matter mixed with
remnant primordial gas (Decressin et al. 2007).
The uncertainties related to the cross-sections strongly
limit the predictive power of the results obtained, and the observational
spread ($\delta$\nafe $\sim 0.3$ for a given \ofe) makes the comparison
not straightforward. However, we note that the ejecta of models with masses
$M \geq 5M_{\odot}$ might account for the oxygen and sodium abundances
detected in the 2nd generation of stars.
{\bf Notice that here we are touching again the problem of the mass budget: if only the
envelopes of stars from 5 up to 6.3\msun\ ---or at most up to 7-8\msun, if we can assume that 
also the superAGBs contribute with similar yields--- can form the self--enriched stars,
the gas contained in this mass range, for reasonable IMFs, is only a few percent of the
total initial cluster mass. As we remarked in the introduction, we must then hypothize
that the initial cluster mass was much larger than the present mass, and that 
preferentially the second generation stars have been lost during the long term
cluster evolution.}

The right panel of Fig.\ref{antic} reports the observed points in the
O-Al plane. Even in this case we report two theoretical lines,
obtained with different choices of the rates of the Mg-Al chain
reactions (the upper line refers to models calculated with the maximum 
allowed values of the cross-section of the proton capture reactions by 
the two heavy magnesium isotopes). In this case the observed trend is 
well reproduced, in particular for the chemistry of the most oxygen poor 
stars, which evidentiate an aluminum enhancement by \alfe$\sim$1. 
In the case of aluminum the comparison is more straightforward, because 
the theoretical uncertainties related to the cross-sections are smaller.
The two squares in the figure have the same meaning as in the left panel.

A welcome result from this investigation is that the lithium content
of the ejecta of the most massive models, those showing the strongest
depletion of oxygen, is A(Li)$\sim 2$ (see teb.\ref{tabchim}), and are
therefore in excellent agreement with the recent analysis of TO stars
in NGC 6752 by Pasquini et al. (2005). Actually, the overall O-Li trend
in the range \ofe$<0.2$ is well reproduced (see their fig.3).

Although the observed range of magnesium abundances is small, so that the 
existence of a real spread is still debated, our models indicate that
a Mg-Al anticorrelation should be present, with a spread 
$\delta$\mgfe $\sim 0.3$dex.

Finally, our models point in favour of a strong depletion of fluorine, 
which is burnt heavily at the very beginning of the AGB phase in all models 
whose initial mass exceeds 4\msun. This is in agreement with a recent
investigation focused on M4 giants (Smith et al. 2005); a robust
confirm of this scenario would be the determination of fluorine abundances
in TO and SGB stars.

\section{Conclusions}
We present updated model for the evolution of AGB stars of 
$3M_{\odot}\leq M \leq 6.3M_{\odot}$, Y=0.24 and Z=0.001.
We test the self-enrichment scenario hypothesis for GCs of intermediate
metallicity, comparing the results of spectroscopic investigations of scarcely
evovlved stars in GCs with the theoretical yields of intermediate mass stars.
This task demanded the computation of a new set of models, calculated with
the latest physical updates concerning the equation of state and the 
conductive opacities, and with an $\alpha-$enhanced mixture, to describe
self-consistently the chemistry of the first generation of stars which form
in the GCs.

\begin{figure*}
\centering{
}
\caption{Left: The observed Oxygen-sodium trend in stars in GCs. The solid lines
indicate the content of the ejecta of our models, obtained with the recommended
cross-sections for the reactions involved in the Ne-Na cycle (lower track), and
with the reactions rates maximizing sodium production (upper lines); the dashed 
track shows the abundances of the ejecta from the models calculated with the
enhanced cross sections for the $\alpha-$captures by $^{22}$Ne. Right: The
observed Oxygen-Aluminum anticorrelation. The two solid lines indicate 
the results from our models, accroding to the choices made for the cross-section
of the Mg-Al reactions. The observed points refer to the following works.
Full triangles: NGC 6752 (Gratton et al. 2001); Open squares: M5 stars with
$V>16$ (Ramirez \& Cohen 2003); Stars: M13 stars with $V>15$ (Cohen \& Melendez 2005);
Open triangles: M3 stars with $V>15$ (Cohen \& Melendez 2005); Full points: 
NGC 6218 stars (Carretta et al. 2006); Open squares: High gravity M3 giants
(Sneden et al. 2004); Full squares: High gravity M13 giants (Sneden et al. 2004)}
      \label{antic}%
\end{figure*}

The high efficiency of the convective model adopted confirms an important result
obtained by this research group regarding the AGB evolution of intermediate mass 
stars, i.e. the possibility of a strong nucleosynthesis at the bottom of the 
external convective zone for all the masses M$>3$\msun; this HBB also favours
a fast increase of the luminosity, a higher mass loss, and therefore reduces
the number of thermal pulses experienced by the star during the AGB phase.
The present investigation indicate that for this metallicity a maximum 
$\tbce=110$MK may be reached for the highest masses; this upper limit, and the 
strong mass loss suffered by the most massive models, limits the extent of the 
nucleosynthesis which may be achieved within the envelope.

The combination of HBB and TDU in the most massive models favours a strong
depletion of oxygen and fluorine, a modest reduction of magnesium, and a large 
production of aluminum. Sodium is also produced, via a delicate compromise
between production by neon burning and destruction by proton capture. It is
confirmed that lithium can be produced at the beginning of the AGB phase
via the Cameron-Fowler mechanism, to be later destroyed due to $^3$He 
consumption within the envelope. The matter ejected by these models is
helium rich, which a maximum enrichment of Y=0.35 found for the 6\msun model.

On the basis of the results of this work, the strongest point in favour of 
the self-enrichment scenario is the oxygen depletion, for which both the 
observations an the theoretical predictions indicate a maximum limit of 
\ofe$=-0.4$; the theoretical oxygen yield is robust, as it turns out to be 
approximately independent of the details of the mass loss description, and 
the relevant cross-sections are known with sufficient accuracy.

The O-Al trend is confirmed by the present investigation, though the extent of
the aluminum enrichment of the ejecta is sensitive to the assumptions 
regarding the cross-sections of the proton capture reactions by the heavy
magnesium isotopes and by $^{26}$Al: note that the largest enrichment, 
\alfe$=1$ is consistent with the aluminum content of the ejecta of the largest 
masses studied here, which are also those mostly depleted in oxygen.
The production of aluminum is associated to magnesium burning, which we
predict to be sligthly depleted, in agreement with the findings of some
research groups.

The sodium yield is the most uncertain, due to the huge uncertainties 
associated to the reactions of the Ne-Na cycle. Our models 
confirm that when the maximum allowed values for the reaction rates of 
the proton capture reaction by $^{22}$Ne nuclei are adopted, the most oxygen 
poor ejecta are also sodium rich, but the exact extent of
the sodium enrichment, and the confirm that a clear anticorrelation exists,
can be hardly fixed with the present cross-sections: the theoretical 
uncertainties related to the sodium content amount to 0.6 dex.

The yield of lithium and fluorine are most sensitive to the mass loss rate
adopted. With our standard choice, our models predict a O-Li trend which is
in excellent agreeent with a recent investigation based on the lithium 
content of TO stars in NGC 6752. The fluorine content is expected to
be extremely poor in any case, the exact abundance being determined by 
the details of the mass loss description.


\end{document}